\documentclass[12pt]{article}
\pdfoutput=1
\usepackage{jhep-mod}
\usepackage{bm}
\usepackage{amssymb,amsmath,amsthm}
\usepackage{mathrsfs} 

\begin{document}
\title{Classical and semi-classical energy conditions\footnote{Draft chapter, on which the related chapter of the book {\it Wormholes, Warp Drives and Energy Conditions} (to be published by Springer) will be based.}}
\author[a]{Prado Martin-Moruno}
\author[b]{\,{\sf and}\, Matt Visser}
\affiliation[a]{Departamento de F\'isica Te\'orica I, Ciudad Universitaria, Universidad Complutense de Madrid,\\
E-28040 Madrid, Spain}
\affiliation[b]{School of Mathematics and  Statistics, Victoria University of Wellington, \\
PO Box 600, Wellington 6140, New Zealand}
\emailAdd{pradomm@ucm.es}
\emailAdd{matt.visser@sms.vuw.ac.nz}
\abstract{
The standard energy conditions of classical general relativity are (mostly) linear in the stress-energy tensor, and have clear physical interpretations in terms of geodesic focussing, but suffer the significant drawback that they are often violated by semi-classical quantum effects. In contrast, it is possible to develop non-standard energy conditions that are intrinsically non-linear in the stress-energy tensor, and which exhibit much better well-controlled behaviour when semi-classical quantum effects are introduced,
at the cost of a less direct applicability to geodesic focussing. In this chapter we will first review the standard energy conditions
 and their various limitations. (Including the connection to the Hawking--Ellis type I, II, III, and IV classification of stress-energy tensors).  We shall then turn to the averaged, nonlinear, and semi-classical energy conditions, and see how much can be done once semi-classical quantum effects are included. 
}
\maketitle

\section{Introduction}
\label{sec:1}

The energy conditions, be they classical, semiclassical, or ``fully quantum'' are at heart purely phenomenological approaches to deciding what form of stress-energy is to be considered ``physically reasonable''. As Einstein put it, the LHS of his field equations $G_{ab} = 8\pi G_N \, T_{ab}$  represents the  ``purity and nobility'' of geometry, while the RHS represents ``dross and uncouth'' matter. If one then makes no assumptions regarding the stress-energy tensor, then the Einstein field equations are simply an empty tautology. It is only once one makes \emph{some} assumptions regarding the stress-energy tensor, (zero for vacuum, ``positive'' in the presence of matter), that solving the Einstein equations becomes closely correlated with empirical reality. The energy conditions are simply various ways in which one can try to implement the idea that the stress-energy tensor be ``positive'' in the presence of matter, and that gravity always attracts.
In this chapter we shall first discuss the standard point-wise energy conditions, and then move on to their generalizations obtained by averaging along causal curves, and some nonlinear variants thereof, and  finally discuss semi-classical variants.

\section{Standard energy conditions}
\label{sec:2}

General relativity, as well as other metric theories of gravity, proposes a way in which the matter content affects the curvature of the spacetime. The theory of gravity itself, however, does not \emph{per se} tell us anything specific about the kind of matter content that we should consider.
If we are interested in extracting general features about the spacetime, independently of the form of the stress-energy tensor of matter fields, we have to consider purely geometric expressions such as the Raychaudhuri equation. This equation describes the focussing of a congruence of timelike curves in a given geometry. Particularized to geodesic motion it can be expressed as~\cite{H&E}
\begin{equation}\label{R}
\frac{{\rm d}\theta}{{\rm d}s}=\omega_{ab}\omega^{ab}-\sigma_{ab}\sigma^{ab}-\frac{1}{3}\theta^2-R_{ab}V^aV^b,
\end{equation}
where $\omega_{ab}$ is the vorticity, $\sigma_{ab}$ is the shear, $\theta$ is the expansion, and $V^a$ is the timelike unit vector tangent to the congruence. 
Note that for a congruence of time-like geodesics with zero vorticity, for which $V_a\propto\nabla_a\Phi(x)$ with $\Phi(x)$ a scalar field \cite{Abreu:2010eb}, the only term on the RHS 
of equation (\ref{R}) that can be positive is the last term. Hence, if $R_{ab}V^aV^b\geq0$, then the expansion decreases and, 
as free-falling observers follow geodesics, observers that only feel the gravitational interaction get closer to each other.
This fact allows one to formulate the following geometric condition:

\noindent\emph{
Time-like convergence condition (TCC): Gravity is always attractive provided
\begin{equation}
R_{ab}V^aV^b\geq0\quad {\sl for\,\, any\,\, timelike\,\, vector\,\,} V^a.
\end{equation}
}
\noindent Note that although the TCC ensures focussing of timelike geodesics the contrary statement is not necessarily true, that is $R_{ab}V^aV^b<0$ does not automatically imply geodesics defocusing \cite{Borde:1987qr}.
As $R_{ab}$ depends on the stress-energy tensor through the equations of the dynamics, the TCC imposes restrictions on the material content that depends on the specific theory of gravity.

Analogously, one can take the equation for the focussing of a congruence of null geodesics with vanishing vorticity, and formulate the null convergence condition\footnote{There have been interesting attempts to derive the NCC from an underlying fundamental framework \cite{Parikh:2014mja,Parikh:2015wae,Parikh:2015ret,Parikh:2016lys}.} (NCC). This is a particular limit of the TCC and requires that $R_{ab}k^ak^b\geq0$ for any null-vector $k^a$.
It should be noted that hypothetical astronomical objects like wormholes would necessarily violate the NCC \cite{Morris:1988cz,Morris:1988tu,Hochberg:1998ha}. 
Furthermore, this condition might also be violated during the early universe if one wishes to avoid the big bang singularity \cite{Borde:1987qr,Borde:1997pp,MolinaParis:1998tx}.

\clearpage

Following a different spirit, one could consider imposing an assumption on the form of the stress-energy tensor based on our daily experience through some energy condition. This can be done restricting attention to a given theory of gravity. Such is the case for the strong energy condition which is formulated as:

\noindent
\emph{
Strong energy condition (SEC): Gravity is always attractive in GR.
\begin{equation}\label{sec} 
\left(T_{ab} -\frac{1}{2}Tg_{ab} \right) V^aV^b\geq0 \quad {\sl for\,\, any\,\, timelike\,\, vector\,\,} V.                                           
\end{equation}
}
\noindent This SEC condition takes into account both the TCC \emph{and the Einstein equations} in its mathematical formulation. Hence, properly speaking, it only makes sense to refer to the SEC in a general relativistic framework.

More generally, by looking at the world around us we can also assert certain properties that any ``reasonable'' material content should satisfy \emph{independently of the theory of gravity} describing the curvature of spacetime. These energy conditions can be used later to study the geometry once the theory of gravity is fixed. The energy conditions commonly considered in the literature are \cite{H&E,Barcelo:2002bv}:

\noindent\emph{
Dominant energy condition (DEC): The energy density measured by any observer is non-negative and  propagates in a causal way.
\begin{eqnarray}\label{dec}
T^a{}_b\,V_a\, V^b\geq0,\qquad 
F^aF_a\leq0.
\end{eqnarray}
}
\noindent Here we have defined the flux 4-vector as $F^a=-T^{a}{}_bV^b$, and it is understood that the inequalities contained in expression (\ref{dec}) have to be satisfied for any timelike vector.

\noindent
\emph{
Weak energy condition (WEC): The energy density measured by any observer has to be non-negative.
\begin{equation}\label{wec}
T^a{}_b\,V_a\, V^b\geq0.
\end{equation}
}
\noindent Therefore, if the WEC is not satisfied, then the DEC has to be violated. However, the WEC and the SEC are, in principle, completely independent.

\noindent
\emph{
Null energy condition (NEC): The SEC and the WEC are satisfied in the limit of null observers.
\begin{equation}\label{nec}
T^a{}_b\,k_a\, k^b\geq0\quad{\sl for\,\, any\,\, null\,\, vector\,\,} k^a.  
\end{equation}
}
\noindent It should be noted that the NCC is completely equivalent to the NEC within the framework of GR. By its very definition if the NEC is violated, then the SEC, on one hand, and the WEC and DEC, on the other hand, cannot be satisfied.

Historically, there was also a trace energy condition (TEC), the assertion that the trace of the stress energy tensor should (in $-+++$ signature) be non-positive. While for several decades in the 50s and 60s this was believed to be a physically reasonable condition, opinion has now shifted, (specifically with the discovery of stiff equations of state for neutron star matter), and the TEC has now largely been abandoned~\cite{Barcelo:2002bv}. (See also reference \cite{Bekenstein:2013ztp} for uses of the less known subdominant trace energy condition, and reference \cite{Hayward:1994cu} for the practically unnoticed quasilocal energy conditions.) This cautionary tale indicates that one should perhaps take the energy conditions as provisional, they are good ways of qualitatively and quantitatively characterizing the level of weirdness one is dealing with, but they may not actually be fundamental physics.

\subsection{The Hawking--Ellis type I --- type IV classification}
\label{sec:2.1}

The Hawking--Ellis classification, (more properly called the Segre classification), of stress-energy tensors is based on the extent to which the ortho-normal components of the stress-energy can be diagonalized by local Lorentz transformations:
\begin{equation}
T^{ab} =   (\hbox{canonical type})^{cd} \; L_c{}^a\; L_d{}^b.
\end{equation} 
Equivalently, one is looking at the Lorentz invariant eigenvalue problem
\begin{equation}
\det\left(T^{ab} - \lambda \eta^{ab}\right) = 0, \quad \hbox{that is}, \qquad 
\det\left(T^a{}_b - \lambda \delta^a{}_b\right) = 0.
\end{equation}
Here as usual $\eta^{ab}={\rm diag}(-1,\,1,\,1,\,1)$. 
In Lorentzian signature $T^a{}_b$ is not symmetric, so diagonalization is trickier than one might naively expect.
Even the usual Jordan decomposition is not particularly useful, 
(since for physical reasons one is interested in partial diagonalization using Lorentz transformations, rather than the more general similarity transformations which could lead to non-diagonal unphysical $T^{ab}$), and so it is more traditional to classify stress-energy tensors in terms of the spacelike/lightlike/timelike nature of the eigenvectors.

\noindent
Based on (a minor variant of) the discussion of Hawking and Ellis the four types (when expressed in an orthonormal basis) are:

\begin{itemize}
\item \underline{\underline{Type I:}} 
\begin{equation}
T^{ab} \sim \left[ \begin{array}{c|ccc} \rho&0&0&0\\ \hline0&p_1&0&0\\0&0&p_2&0\\0&0&0&p_3 \end{array} \right].
\end{equation}
This is the generic case, there is one timelike, and 3 spacelike eigenvectors.
\begin{equation}
T^{ab} = \rho \; u^a u^b + p_1 \; n_1^a n_1^b + p_2 \; n_2^a n_2^b + p_3 \; n_3^a n_3^b,
\end{equation}
where $u^a=(-1;\,0;\,0; \,0)$,\, $n_1^a=(0;\,1;\,0;\,0)$,\, $n_2^a=(0;\,0;\,1;\,0)$, and $n_3^a=(0;\,0;\,0;\,1)$.

\enlargethispage{35pt}
The Lorentz invariant eigenvalues of $T^a{}_b$ are $\{-\rho,\,p_1,\,p_2,\,p_3\}$. 
Many classical stress-energy tensors, (for instance, perfect fluids, massive scalar fields, non-null electromagnetic fields), are of this type~I form. Similarly many semi-classical stress-energy tensors are of this type~I form. 

For a type I stress tensor the classical energy conditions can be summarized (in terms of necessary and sufficient constraints) as:
\begin{center}
\begin{tabular}{||c||c|c||}
\hline
\vphantom{\Big{|}} NEC  & $\;\rho+p_i \geq 0\;$& --- \\
\vphantom{\Big{|}} WEC  & $\rho+p_i \geq 0$& $\rho \geq 0$\\
\vphantom{\Big{|}} SEC  & $\rho+p_i \geq 0$& $\;\rho+\sum p_i  \geq 0\;$\\
\vphantom{\Big{|}} DEC  & $|p_i|\leq \rho$ &  $\rho \geq 0$\\
\hline
\end{tabular}
\end{center}

\item  \underline{\underline{Type II:}}
\begin{equation}
T^{ab} \sim \left[ \begin{array}{cc|cc} \mu+f&f&0&0\\ f&-\mu+f &0&0\\ \hline0&0&p_2&0\\0&0&0&p_3 \end{array} \right] 
\end{equation}
This corresponds to one double-null eigenvector $k^a = (1;\,1; \,0;\, 0)$, so that:
\begin{equation}
T^{ab} = + f \; k^a k^b - \mu \; \eta_2^{ab} + p_2 \; n_2^a n_2^b + p_3 \; n_3^a n_3^b,
\end{equation}
with $\eta_2^{ab}={\rm diag}(-1,\,1,\,0,\,0)$.\footnote{For type II Hawking and Ellis choose to set $f\to\pm1$, which we find unhelpful.}
The Lorentz invariant eigenvalues of $T^a{}_b$ are now $\{-\mu,-\mu,\,p_2,\,p_3\}$. 
Classically, the chief physically significant observed occurrence of type II stress-energy is when $\mu=p_i=0$, in which case one has
\begin{equation}
T^{ab} = f \; k^a\,  k^b.
\end{equation}
This corresponds to classical radiation or null dust. 

One could also set $p_i=-\mu$ in which case
\begin{equation}
T^{ab} = -\mu \eta^{ab} + f \;k^a \,k^b.
\end{equation}
This corresponds to a superposition of cosmological constant and classical radiation or null dust. 
Finally, if one sets  $p_i=+\mu$ then one has
\begin{equation}
T^{ab} \sim \left[ \begin{array}{cc|cc} \mu&0&0&0\\ 0&-\mu &0&0\\ \hline0&0&\mu&0\\0&0&0&\mu \end{array} \right] 
+  f \;k^a \,k^b.
\end{equation}
This corresponds to a classical electric (or magnetic) field of energy density $\mu$ aligned with classical radiation or null dust.

\clearpage
For a type II stress tensor the classical energy conditions can be summarized as:
\begin{center}
\begin{tabular}{||c||c|c|c||}
\hline
\vphantom{\Big{|}} NEC  & $\;f> 0\;$& $\mu+p_i \geq 0$ & --- \\
\vphantom{\Big{|}} WEC  &$f> 0$& $\mu+p_i \geq 0$& $\mu\geq 0$\\
\vphantom{\Big{|}} SEC  &$f> 0$& $\mu+p_i \geq 0$&$\;\sum p_i \geq 0\;$\\
\vphantom{\Big{|}} DEC  & $f> 0$&  $\;|p_i|\leq \mu\;$ &  $\mu\geq 0$\\
\hline
\end{tabular}
\end{center}

\item  \underline{\underline{Type III:}}
\begin{equation}
T^{ab} \sim \left[ \begin{array}{ccc|c} \rho&0&f&0\\ 
0&-\rho&f&0\\f&f&-\rho&0\\ \hline0&0&0&p \end{array} \right] 
\end{equation}
This corresponds to one triple-null eigenvector $k^a = (1;\,1;\, 0; \,0)$. 
\begin{equation}
T^{ab} = -\rho \;\eta_3^{ab} + f \; (n_2^a k^b +k^a n_2^b) + p \; n_3^a n_3^b,
\end{equation}
with $\eta_3^{ab}={\rm diag}(-1,\,1,\,1,\,0)$.\footnote{For type III Hawking and Ellis choose to set $f\to1$, which we find unhelpful.}
The Lorentz invariant eigenvalues of $T^a{}_b$ are now $\{-\rho,-\rho,-\rho,\,p\}$.
Classically, type III does not seem to occur in nature --- the reason for which becomes ``obvious'' once we look at the standard energy conditions.
Even semi-classically this type III form is  forbidden (for instance) by either spherical symmetry or planar symmetry --- we know of no physical situation where type III stress-energy tensors arise.

For a type III stress tensor the NEC cannot be satisfied for $f\neq0$, and so 
the only way that any of the standard energy conditions can be satisfied for a type III stress-energy tensor is if $f=0$, in which case it reduces to type I. So if one ever encounters a nontrivial type III stress-energy tensor, one immediately knows that all of the standard energy conditions are violated. 

(This of course implies that any classical type III stress tensor, since it would have to violate all the energy conditions, would likely be grossly unphysical --- so it is not too surprising that no classical examples of type III stress-energy have been encountered.)

\item  \underline{\underline{Type IV:}}
\begin{equation}
T^{ab} \sim \left[ \begin{array}{cc|cc} \rho&f&0&0\\ f&-\rho&0&0\\ \hline 0&0&p_1&0\\ 0&0&0&p_2 \end{array} \right] 
\end{equation}
This corresponds to no timelike or null eigenvectors.  The Lorentz invariant eigenvalues of $T^a{}_b$ are now 
$\{- \rho+if, -\rho-if, p_1, p_2\}$.~\footnote{For type IV Hawking and Ellis choose 
\begin{equation}
T^{ab} \sim \left[ \begin{array}{cc|cc} 0&f&0&0\\ f&-\mu&0&0\\ \hline 0&0&p_1&0\\ 0&0&0&p_2 \end{array} \right] 
\qquad (\mu^2 < 4 f^2).
\end{equation}
We have found the version presented in the text to be more useful.}

Classically, type IV does not seem to occur in nature. 
Again, a necessary condition for the NEC to be satisfied for a type IV stress tensor is $f=0$.
The only way that any of the standard energy conditions can be satisfied for a type IV stress-energy tensor is if $f=0$, in which case it reduces to type I. So whenever one encounters a nontrivial type IV stress-energy tensor, one immediately knows that all of the standard energy conditions are violated. (Again, this of course implies that any classical type IV stress tensor, since it would have to violate all the energy conditions, would likely be grossly unphysical --- so it is not too surprising that no classical examples of type IV stress-energy have been encountered.)

Semi-classically, however, vacuum polarization effects do often generate type IV stress energy \cite{Roman:1986tp}.
The renormalized stress-energy tensor surrounding an evaporating (Unruh vacuum) black hole will be type IV in the asymptotic region \cite{Martin-Moruno:2013wfa}, (with spherical symmetry enforcing $p_1=p_2$). As it will be type I close to the horizon, there will be a finite radius at which it will be type II.

\item  \underline{\underline{Euclidean signature:}}

Since we have seen this point cause some confusion, it is worth pointing out that when working in Euclidean signature (in an orthonormal basis) the 
eigenvalue problem simplifies to the standard case
\begin{equation}
\det\left(T^{ab} - \lambda \delta^{ab}\right) = 0.
\end{equation}
Since $T^{ab}$ is symmetric it can be diagonalized by ordinary orthogonal 4-matrices. The symmetry group is now $O(4)$,  not $O(3,1)$.  So in Euclidean signature \emph{all} stress energy tensors are of type I.

The other thing that happens in Euclidean signature us that, because one no longer has the timelike/lightlike/spacelike distinction, both the NEC and DEC simply cannot be formulated. In fact it is best to abandon even the WEC, and concentrate attention on a Euclidean variant of the SEC:
\begin{itemize}
\item \emph{Euclidean Ricci convergence condition: \,\, $R_{ab} V^a V^b \geq 0$ for all vectors. \vphantom{\Big{|}}}
\item \emph{Euclidean SEC: \,\, $(T_{ab}-{1\over2}\,T\, \eta^{ab}) V^a V^b \geq 0$ for all vectors.}
\end{itemize}

\item  \underline{\underline{Classifying the Ricci and Einstein tensors:}} 

Instead of applying the Hawking--Ellis type I --- type IV classification to the stress-energy tensor, one could choose to apply these ideas directly to the Ricci tensor (or Einstein tensor). The type~I --- type IV classification would now become statements regarding the number of spacelike/null/eigenvectors of the Ricci/Einstein tensor, and the associated eigenvalues would have a direct interpretation in terms of curvature, (rather than densities, fluxes, and stresses). Formulated in this way, this directly geometrical approach would then lead to constraints on spacetime curvature that would be independent of the specific gravity theory one works with.

\end{itemize}

\subsection{Alternative canonical form}
\label{sec:2.2}

Sometimes it is useful to side-step the Hawking--Ellis classification and, using only rotations, reduce an arbitrary stress-energy tensor to the alternative canonical form
\begin{equation}
T^{ab} \sim \left[ \begin{array}{c|ccc} \rho&f_1&f_2&f_3\\ \hline f_1&p_1&0&0\\  f_2&0&p_2&0\\ f_3&0&0&p_3 \end{array} \right] 
\end{equation}
The advantage of this alternative canonical form is that it covers all of the Hawking--Ellis types simultaneously. The disadvantage is that it easily provides us with necessary conditions, but does not easily provide us with sufficient conditions for generic stress tensors. For example, enforcing the NEC requires
\begin{equation}
\rho\pm2 \sum_i\beta_if_i+\sum_i\beta_i^2p_i\geq0,\quad{\rm with}\quad\sum_i\beta_i^2=1.
\end{equation}
By summing over $\pm\beta$ one obtains
\begin{equation}
\rho+\sum_i\beta_i^2p_i\geq0,\quad{\rm with}\quad\sum_i\beta_i^2=1,
\end{equation}
which in turn implies
\begin{equation}
\hbox{NEC:}  \qquad\qquad\qquad \left|  \sum_i\beta_if_i \right| \leq 
{1\over2}\left(\rho+\sum_i\beta_i^2p_i\right) \geq0,\quad{\rm with}\quad\sum_i\beta_i^2=1. 
\end{equation}
These are necessary and sufficient conditions for the NEC to hold --- the quadratic occurrence of the $\beta_i$ makes it difficult to simplify these further. Similarly  for the WEC one finds
\begin{equation}
\hbox{WEC:}  \qquad\qquad\qquad \left|  \sum_i\beta_if_i \right| \leq 
{1\over2}\left(\rho+\sum_i\beta_i^2p_i\right) \geq0,\quad{\rm with}\quad\sum_i\beta_i^2\leq1. 
\end{equation}
For the SEC one obtains
\begin{equation}
\hbox{SEC:}  \qquad \left|  \sum_i\beta_if_i \right| \leq 
{1\over4} \left( \left(\rho+\sum_i p_i\right) +\sum_i\beta_i^2\left(\rho+p_i-\sum_{j\neq i} p_j\right)\right)\geq0,
\end{equation}
again with $\sum_i\beta_i^2 \leq1$. 
Finally, the DEC is somewhat messier. The DEC amounts to the WEC \emph{plus} the relatively complicated constraint
\begin{equation}
\left(\rho + \sum_i \beta_i f_i\right)^2 - \sum_i (f_i + \beta_i p_i)^2  \geq 0, \quad{\rm with}\quad\sum_i\beta_i^2\leq1. 
\end{equation}
This can be slightly massaged to yield
\begin{equation}
 \left|  \sum_i\beta_if_i(\rho-p_i)  \right| \leq 
{1\over2}\left(\rho^2+ \left(\sum_i\beta_i f_i\right)^2 - \sum_i (f_i^2 + \beta_i^2 p_i^2) \right) \geq0,
\end{equation}
again with $\sum_i\beta_i^2 \leq1$. 
Unfortunately this seems to be the best that can generically be done for the DEC in this alternative canonical form.

\subsection{Limitations of the standard energy conditions}
\label{sec:2.3}

For practical computations the energy conditions are most often rephrased in terms of the parameters appearing in the type I --- type IV classification, as presented in section \ref{sec:2.1}, with type I stress-energy being the most prominent. 
(The other types are often quietly neglected.)
When expressed in this form, the standard energy conditions can easily be compared with both classical and with published semi-classical estimates of the renormalized expectation value of the quantum stress-energy tensor. 

\clearpage
Beyond the strong physical intuition in which the energy conditions are based, there are deep theoretical reasons why one should wish the energy conditions to be satisfied. It is known that violations of the NEC, which is the weaker energy condition we have presented for the moment, may signal the presence of instabilities and superluminal propagation\footnote{Note that stable violations of the NEC are now known to be possible when the field has a non-canonical kinetic term \cite{Nicolis:2009qm,Deffayet:2010qz,Kobayashi:2010cm}.} \cite{Dubovsky:2005xd,Visser:1999fe,Visser:1998ua,Lobo:2002zf}. 
Moreover, stress-energy tensors violating the NEC can support hypothetical configurations as wormholes  \cite{Morris:1988tu} and warp drives  \cite{Alcubierre:1994tu} in the framework of general relativity\footnote{See references 
\cite{Visser:1989kh,Visser:1989kg,Cramer:1994qj,Ori:1994phl,Kar:1994tz,Hochberg:1997wp,Visser:1997yn,Hochberg:1998ii,Hochberg:1998vm,Hochberg:1998qw,Barcelo:2000ta,Dadhich:2001fu,Visser:2002ua,Lemos:2003jb,Lobo:2003xd,Lobo:2004rp,Roman:2004xm,Lobo:2004an,Lobo:2005us,Lobo:2005vc,Lobo:2007zb,MartinMoruno:2009iu,Visser:2015mur} 
for interesting research along those lines.}, which may lead to pathological situations. 
In addition, the energy conditions are central to proving the positivity of mass \cite{Bekenstein:1975wj} and the singularity theorems \cite{H&E,Borde:1987qr}.
In this context, although one might think that violations of the NEC might always allow us to get rid of uncomfortable singularities, this is not the case \cite{Cattoen:2005dx,Cattoen:2006yh}, and they can instead introduce new kinds of cosmic singularities instead \cite{Starobinsky:1999yw,Caldwell:2003vq,Yurov:2006we,BouhmadiLopez:2006fu,BouhmadiLopez:2007qb}.
Nevertheless, it is already well-known that the energy conditions are violated in realistic situations, violations of these energy conditions in the presence of semi-classical effects being very common.
Among many situations where one or more of the energy conditions are violated we mention:
\begin{itemize}
\item SEC has to be violated in cosmological scenarios during the early phase of inflationary expansion and at the present time \cite{Visser:1997au,Visser:1997qk,Visser:1997tq,Visser:1999de,Cattoen:2007jr}.
\item It is easy to find violations of the SEC associated even with quite normal matter \cite{Kandrup:1992xw,Rose:1987bc}.
\item Non-minimally coupled scalar fields can easily violate all the energy conditions even classically  \cite{Barcelo:2000zf}.\\(However, those fields may be more naturally interpreted as modifications of general relativity in some cases.)
\item The Casimir vacuum violates all the mentioned energy conditions \cite{Roman:1986tp}.
\item Numerically estimated renormalized vacuum expectation values of test-field quantum stress tensors surrounding Schwarzschild and other black holes \cite{Visser:1997gf}. \\(EC violations occur, to one extent or another, for all of the standard quantum vacuum states: Boulware \cite{Visser:1996iv}, Hartle--Hawking \cite{Visser:1996iw}, and Unruh \cite{Visser:1997sd}.)
\item  Explicitly calculable renormalized vacuum expectation values of test-field quantum stress tensors in 1+1 dimensional QFTs \cite{Visser:1996ix}.\\(Again, energy condition violations occur to one extent or another, for all of the standard quantum vacuum states: Boulware, Hartle--Hawking, and Unruh.) 
\end{itemize}

Finally, it should be emphasized that the DEC, WEC, and NEC impose restrictions on the form of the stress-energy tensor of the matter fields independently of the theory of gravity. However, when these conditions are used to extract general features of the spacetime, (such as, for example, the occurrence of singularities), one considers a particular theory of gravitation. Therefore, although the existence of wormholes and bouncing cosmologies required violations of the NEC in general relativity, in modified gravity theories the NCC could be violated by material content satisfying the NEC. In that case, it is possible to define an effective stress-energy tensor putting together the modifications with respect to general relativity that violates the NEC \cite{Barcelo:2000zf,Bellucci:2001cc,Baccetti:2012re,Capozziello:2013vna,Capozziello:2014bqa,Rubakov:2014jja}. As this effective tensor is not associated with matter, there is no reason to require that it had some {\it a priori} characteristics. Hence, one may obtain wormhole solutions supported by matter with ``positive'' energy \cite{Lobo:2009ip,MontelongoGarcia:2010xd,Bohmer:2011si}, as in some of the examples reviewed in the first part of this book.

\section{Averaged energy conditions}
\label{sec:3}
\noindent
The key to the averaged energy conditions is simply to integrate the Raychaudhuri equation along a timelike geodesic
\begin{equation}\label{R}
\theta_f - \theta_i=\int_i^f \left\{ \omega_{ab}\omega^{ab}-\sigma_{ab}\sigma^{ab}-\frac{1}{3}\theta^2-R_{ab}V^aV^b\right\} d s.
\end{equation}
Assuming zero vorticity, and noting that both $\sigma_{ab}\sigma^{ab}\geq 0$ and $\theta^2\geq 0$ we have
\begin{equation}\label{R}
\theta_f \leq  \theta_i - \int_i^f R_{ab}V^aV^b d s.
\end{equation}
That is, a constraint on the \emph{integrated} timelike convergence condition, $\int R_{ab}V^aV^b d s$, is sufficient to control the convergence of timelike geodesics. In applications, the integral generally runs from some base point along a timelike geodesic into the infinite future. We formulate the ATCC (averaged timelike convergence condition):

\noindent\emph{
Averaged time-like convergence condition (ATCC): Gravity attracts provided
\begin{equation}
\int_0^\infty R_{ab}V^aV^b ds\geq 0\quad \hbox{along any timelike geodesic}.
\end{equation}
}
\noindent
In standard general relativity the Einstein equations allow one to reformulate this as

\noindent \emph{
Averaged strong condition (ASEC): Gravity in GR attracts on average.
\begin{equation}
\int_0^\infty \left(T_{ab}-{1\over2} T g_{ab}\right) V^aV^b ds\geq 0\quad \hbox{along any timelike geodesic}.
\end{equation}
}
\noindent
Similar logic can be applied to null geodesic congruences and the Raychaudhuri equation along null geodesics to deduce

\noindent\emph{
Averaged null convergence condition (ANCC): Gravity attracts provided
\begin{equation}
\int_0^\infty R_{ab}k^ak^b d\lambda \geq 0\quad \hbox{along any null geodesic}.
\end{equation}
}
\noindent
In standard general relativity the Einstein equations allow one to reformulate this as

\noindent\emph{
Averaged null energy condition (ANEC): A necessary requirement for gravity to attract in GR on average is
\begin{equation}
\int_0^\infty T_{ab} \; k^ak^b \; d\lambda\geq 0\quad \hbox{along any null geodesic}.
\end{equation}
}
\noindent
Note that the integral must be performed using an affine parameter --- with arbitrary parameterization these conditions would be vacuous.  
This ANEC has attracted considerable attention over the years\footnote{Interesting studies include (but are not limited to) references \cite{Klinkhammer:1991ki,Ford:1994bj,Yurtsever:1994wc,Ford:1995gb,Flanagan:1996gw,Fewster:2002ne,Graham:2007va}.} to prove singularity theorems \cite{Roman:1988vv,Fewster:2010gm} and is for instance the basis of the \emph{topological censorship theorem} \cite{Friedman:1993ty,Friedman:2008dh} forbidding a large class of traversable wormholes. 
However, note that even the ANEC can be violated by conformal anomalies \cite{Visser:1994jb}, so to have any hope of a truly general derivation of the ANEC from more fundamental principles, one would need to enforce some form of conformal anomaly cancellation.

\section{Nonlinear energy conditions}
\label{sec:4}

The idea behind nonlinear energy conditions is inspired to some extent by the qualitative difference between the NEC/WEC/SEC and the DEC. The NEC/WEC/SEC are strictly linear --- any sum of stress-energy tensors satisfying these conditions will also satisfy the same condition. This linearity fails however for the DEC, which can be rephrased as the two conditions
\begin{equation}
T_{ab} \, V^a V^b \geq 0; \qquad (-T_{ac}\eta^{cd} T_{db}) V^a V^b \geq 0;  \qquad \forall \quad\hbox{timelike}\quad V. 
\end{equation}
The second quadratic condition is imply the statement that the flux $F$ be non spacelike. We can further combine these into one quadratic condition
\begin{equation}
(-[T_{ac}-\epsilon\eta_{ac}]\eta^{cd} [T_{db}-\epsilon\eta_{db}]) V^a V^b \geq \epsilon^2;  \quad 
\forall \epsilon>0; \quad \forall \quad\hbox{timelike}\quad V. 
\end{equation}

\clearpage
\noindent
That is, the DEC (suitably rephrased) is a nonlinear quadratic constraint on the stress-energy, and it is this version of the DEC that can naturally be integrated along timelike geodesics to develop an ADEC.  Furthermore this observation opens up the possibility that other nonlinear constraints on the stress-energy might be physically interesting.  Among such possibilities, we mention the flux energy condition \cite{Abreu:2011fr,Martin-Moruno:2013sfa}, determinant energy condition, and trace-of-square energy condition \cite{Martin-Moruno:2013wfa}.

\noindent\emph{
Flux energy condition (FEC): The energy density measured by any observer propagates in a causal way.
\begin{equation}
F^aF_a\leq0.
\end{equation}
}
\noindent 
It has to be emphasized that the FEC does not assume anything about the \emph{sign} of the energy density. Moreover, by its very definition the DEC is just the combination of the WEC and the FEC.
For the different types of stress energy tensor we have
\begin{center}
\begin{tabular}{||c||c|c||}
\hline
\vphantom{\Big{|}} Type I  & $\;\rho^2\geq p_i^2\;$& --- \\
\vphantom{\Big{|}} Type II  &$\;\mu^2\geq p_i^2\;$  & $\mu f\geq0$\\
\hline
\end{tabular}
\end{center}
The FEC cannot be satisfied for genuine (that is $f \neq 0$) types III and IV.
The FEC is a weakening of the DEC, and is equivalent to the single quadratic constraint
\begin{equation}
 (T_{ac}\eta^{cd} T_{db}) V^a V^b \leq 0;  \qquad \forall \quad\hbox{timelike}\quad V. 
\end{equation}
Once phrased in this way, it becomes clear how to formulate an averaged version of the FEC.

\noindent\emph{
Averaged flux energy condition (AFEC): 
\begin{equation}
\int_0^\infty  (T_{ac}\eta^{cd} T_{db}) V^a V^b ds\leq 0; \quad \hbox{along all timelike geodesics}.
\end{equation}
}

On the other hand, among other non-linear combinations of the stress-energy tensor,  we might consider the following energy conditions:

\noindent\emph{
Determinant energy condition (DETEC): The determinant of the stress-energy tensor is nonnegative.
\begin{equation}
\det \left( T^{ab}\right) \geq 0.\qquad
\end{equation}
}
\noindent In terms of the type decomposition we have
\begin{center}
\begin{tabular}{||c||c||}
\hline
\vphantom{\Big{|}} Type I  & $\rho p_1p_2p_3\geq0$ \\
\vphantom{\Big{|}} Type II  &$-\mu^2p_1p_2\geq 0$ \\
\vphantom{\Big{|}} Type III  &$\rho^3p\geq 0$\\
\vphantom{\Big{|}} Type IV  & $-\left(\rho^2+f^2\right)p_1p_2\geq 0$\\
\hline
\vphantom{\Big{|}} \,\,Alternative\,\,  & $\;\rho p_1p_2p_3  - f_1^2 p_2p_3 - f_2^2 p_3 p_1 - f_3^2 p_2p_2 \geq 0\;$\\
\hline
\end{tabular}
\end{center}

\noindent\emph{
Trace-of-square energy condition (TOSEC): The trace of the squared stress-energy tensor is nonnegative.
\begin{equation}
T^{ab}\, T_{ab} \geq 0.
\end{equation}
}
\noindent In terms of the type decomposition we have
\begin{center}
\begin{tabular}{||c||c||}
\hline
\vphantom{\Big{|}} Type I  & $\rho^2+\sum p_i^2\geq0$ \\
\vphantom{\Big{|}} Type II  &$2\mu^2+\sum p_i^2\geq0$\\
\vphantom{\Big{|}} Type III  &$3\rho^2+p^2\geq 0$\\
\vphantom{\Big{|}} Type IV & $\;2\left(\rho^2-f^2\right)+\sum p_i^2\geq 0\;$ \\
\hline
\vphantom{\Big{|}} \,\,Alternative\,\,  & $\; \sum f_i^2 \leq {1\over2}(\rho^2 + \sum_i p_i^2) \;$\\
\hline
\end{tabular}
\end{center}
Since the TOSEC is basically a sum of squares, it is only in Lorentzian signature that it can possibly be nontrivial --- and even in Lorentzian signature only a type IV stress-energy tensor can violate the TOSEC.

\noindent
Some comments about the fulfilment of these conditions are in order \cite{Martin-Moruno:2013wfa}:
\begin{itemize}
\item DETEC is violated in cosmological scenarios during the early inflationary phase and at the present time in the framework of GR.
\item FEC is violated by the Casimir vacuum, but it can be satisfied by the renormalized stress-energy tensor of the Boulware vacuum of a Schwarzschild spacetime.
\item TOSEC can be violated by the Unruh vacuum of a Schwarzschild spacetime.
\end{itemize}

\section{Semi-classical energy conditions}
\label{subsec:5}

We now note that when considering semi-classical effects the ECs may be violated, although the relevant inequalities are typically not satisfied only by a \emph{small amount}. This fact led us to formulate the semi-classical energy conditions in a preliminary and somewhat vague way, to allow them to quantify the violation of the classical ECs, before considering a particular more specific formulation \cite{Martin-Moruno:2013wfa,Martin-Moruno:2013sfa}.

\noindent\emph{
Quantum WEC (QWEC): The energy density measured by any observer should not be {excessively negative}. 
}\\
\noindent The QWEC, therefore, allows the energy density to be negative in semi-classical situations but its value has to be bounded from below. Now, assuming that this bound depends on characteristics of the system, such as the number of fields $N$, the system 4-velocity $U^a$, and a characteristic distance $L$, we can formulate the QWEC as

\begin{equation}
T^{a}{}_b\, V_a V^b \geq -\zeta\;{\hbar N\over L^4}\; (U_a\,V^a)^2.
\end{equation}
\noindent Here $\zeta$ is a parameter of order unity. 
For the QWEC we have:
\begin{center}
\begin{tabular}{||c||c|c|c||}
\hline
\vphantom{\Big{|}} Type I  & $\;\rho\geq-\zeta\hbar N/L^4\;$ & $\;\rho+ p_i\geq-\zeta\hbar N/L^4\;$& ---  \\
\vphantom{\Big{|}} Type II  & $\mu\geq-\zeta\hbar N/L^4$ & $p_i\geq-\zeta\hbar N/L^4$ & $f\geq-\zeta\hbar N/L^4$\\
\vphantom{\Big{|}} \,Type III\,  &$\rho\geq-\zeta\hbar N/L^4$& $p\geq-\zeta\hbar N/L^4$ & $|f|\leq\zeta\hbar N/L^4$ \\
\vphantom{\Big{|}} Type IV & $\rho\geq-\zeta\hbar N/L^4$& $p_i\geq-\zeta\hbar N/L^4$ & $\;|f|\leq\zeta\hbar N/L^4\;$ \\
\hline
\end{tabular}
\end{center}

\noindent
In a similar way, other quantum (semi-classical) energy conditions have been formulated.

\noindent\emph{
Quantum FEC (QFEC): The energy density measured by any observer either propagates in a causal way, or does not propagate {too badly} in an acausal way. 
}\\
\noindent So, the QFEC allows the flux 4-vector to be (somewhat) space-like in semi-classical situations but its norm has to be bounded from above. It can be written as

\begin{equation}
F^aF_a \leq \zeta \;\left(\hbar\, N\over L^4\right)^2\; (U_a\,V^a)^2.
\end{equation}
\noindent Here $\zeta$ is again a parameter of order unity. For the QFEC we have:
\begin{center}
\begin{tabular}{||c||c|c|c||}
\hline
\vphantom{\Big{|}} Type I  & $\rho^2-p_i^2\geq-\zeta\left(\hbar N/L^4\right)^2$ & --- & --- \\
\vphantom{\Big{|}} Type II  & $\mu^2-p_i^2\geq-\zeta\left(\hbar N/L^4\right)^2$& $\mu f\geq-\zeta\left(\hbar N/L^4\right)^2$& ---\\
\vphantom{\Big{|}} \,Type III \, &$\rho^2-p^2\geq-\zeta\left(\hbar N/L^4\right)^2$ & $|\rho f|\leq\zeta\left(\hbar N/L^4\right)^2$ & $|f|\leq\zeta\hbar N/L^4$ \\
\vphantom{\Big{|}} Type IV & $\;\rho^2-p_i^2\geq-\zeta\left(\hbar N/L^4\right)^2\;$  & $\;|\rho f|\leq\zeta\left(\hbar N/L^4\right)^2\;$  & $\;|f|\leq\zeta\hbar N/L^4\;$ \\
\hline
\end{tabular}
\end{center}

Analogously with the classical energy conditions, the quantum DEC (QDEC) would be satisfied in situations where both the QWEC and QFEC are fulfilled. That is, the QDEC states the following

\noindent\emph{
Quantum DEC (QDEC): The energy density measured by any observer should not be {excessively negative}, and it either propagates in a causal way or does not propagate {too badly} in an acausal way. 
}

\noindent\emph{
Specifically
\begin{eqnarray}
T^{a}{}_b\, V_a V^b \geq -\zeta_W\,{\hbar N\over L^4}\; (U_a\,V^a)^2,\\
F^aF_a \leq \zeta_F \;\left(\hbar\, N\over L^4\right)^2\; (U_a\,V^a)^2.
\end{eqnarray}
}
\noindent In order to get the relevant inequalities for the QDEC to be satisfied for types I --- IV stress-energy tensors, one needs just to combine the inequalities of the QWEC and QFEC tables, taking into account that the two $\zeta$'s should be kept distinct.\\
(For a recent rediscovery of a quantum energy condition along the lines investigated in references \cite{Martin-Moruno:2013wfa,Martin-Moruno:2013sfa,Bouhmadi-Lopez:2014gza} and presented here see reference \cite{Bousso:2015wca}. For non-local quantum energy inequalities see references \cite{Fewster:2002dp,Fewster:2005rp}.)

\noindent
Regarding the semi-classical quantum energy conditions it should be noted that:
\begin{itemize}
\item QECs are, of course, satisfied in classical situations where their classical counterparts are satisfied.
\item The Casimir vacuum satisfies the QECs presented here \cite{Martin-Moruno:2013wfa}.
\item QFEC can be satisfied by the Boulware and Unruh vacuum even in situations where the QWEC is violated \cite{Martin-Moruno:2013sfa}.
\item The QECs can be used as a way of quantifying/minimizing  the violation of the ECs.
\end{itemize}
Concerning the last point it should be clarified we are referring to quantifying the degree of violation of an energy condition for a given amount of matter \cite{Bouhmadi-Lopez:2014gza,Bouhmadi-Lopez:2014cca,Martin-Moruno:2014hfa,Albarran:2015cda}, and not to minimizing the quantity of matter violating the energy condition \cite{Visser:2003yf,Kar:2004hc,Garcia:2011aa}.

\section{Discussion}
\label{sec:6}

As we have seen, the energy conditions (both classical and semi-classical) are numerous and varied, and depending on the context can give one rather distinct flavours of both qualitative and quantitative information --- either concerning the matter content, or concerning the (attractive) nature of gravity.  Variations on the original classical point-wise energy conditions are still under development and investigation, and the status of the energy conditions as fundamental physics should still be considered provisional. 

There is no strict ordering on the set of energy conditions, at best a partial ordering, so there is no strictly ``weakest" energy condition. Nonetheless, perhaps the weakest of the \emph{usual} energy conditions (in terms of the constraint imposed on the stress energy) is the ANEC, which makes it the strongest energy condition in terms of proving theorems. Attempts at developing a general proof of the ANEC are ongoing --- part of the question is exactly what one might mean by ``proving an energy condition'' from more fundamental principles.
It should be noted, however, that even the ANEC can be violated by conformal anomalies, so to have any hope of a truly general derivation of the ANEC from more fundamental principles, one would need to enforce some form of conformal anomaly cancellation.

On the other hand, it should be emphasized that FEC is completely independent of the NEC. Hence, there could be in principle realistic situations where the stress-energy tensor violates the ANEC but satisfies the FEC. Although, as we have reviewed, the FEC can be violated by semi-classical effects, its quantum counterpart is satisfied by the renormalized stress-energy tensor of quantum vacuum states.

An alternative approach is to formulate curvature conditions directly on the Ricci or Einstein tensors, and then use global analysis, (based, for instance, on the Raychaudhuri equation or its generalized variants) to extract information regarding curvature singularities and/or the Weyl tensor.

\section*{Acknowledgements}
PMM acknowledges financial support from the Spanish Ministry of Economy and Competitiveness through the postdoctoral training contract FPDI-2013-16161, and through the project FIS2014-52837-P. 
MV acknowledges financial support via the Marsden Fund administered by the Royal Society of New Zealand.



\end{document}